\documentclass[a4paper,11pt]{article}
\usepackage{amsmath, amssymb}
\usepackage{lineno}
\usepackage{PRIMEarxiv}
\usepackage{subfigure}
\usepackage{float}
\usepackage{textgreek}
\usepackage{ulem} 
\usepackage[utf8]{inputenc} 
\usepackage[T1]{fontenc}    
\usepackage{hyperref}       
\usepackage{url}            
\usepackage{booktabs}       
\usepackage{amsfonts}       
\usepackage{nicefrac}       
\usepackage{microtype}      
\usepackage{lipsum}
\usepackage{fancyhdr}       
\usepackage{graphicx}       
\usepackage{color}
\usepackage{xcolor}
\usepackage{notes2bib}

\title{High speed, high power 2D beam steering 
for mitigation of optomechanical parametric instability in gravitational wave detectors }

\author{Thomas Harder, Margherita Turconi, Rémi Soulard, Walid Chaibi \\ Université Côte d'Azur, OCA, CNRS, ARTEMIS, Boulevard de l'Observatoire F-06304 Nice, France}

\begin{document}
\maketitle
%
%



\begin{abstract}
In this paper we propose a novel strategy to control optomechanical parametric instability (PI) in gravitational wave (GW) detectors, based on radiation pressure. The fast deflection of a high power beam is the key element of our approach.
We built a 2D deflection system based on a pair of acousto-optic modulators (AOMs) that combines high rapidity and large scan range. As fast frequency switching configurable AOM driver 
we used an Universal Software Radio Peripheral (USRP) combined with a high performance personal computer (PC). In this way we demonstrate a 2D beam steering system with flat efficiency over the whole scan range and with a transition time of 50 ns between two arbitrary consecutive deflection positions for a beam power of 3.6 W. 
\end{abstract}

\section{Introduction}

The ground-based GW detectors \cite{ligo,virgo,kagra} are km-scale laser interferometers with Fabry-Perot cavities in their arms that consist of high mechanical quality-factor mirrors. That kind of mirrors combined with the high optical power circulating in the cavities gives rise to a nonlinear optomechanical phenomenon, called parametric instability (PI) which causes the dysfunction of the detector \cite{PI_obs}.   

PI involves three modes: the fundamental optical cavity mode, a higher order optical mode and a mechanical eigenmode of one cavity mirror. A part of the fundamental optical mode gets scattered on a thermally excited cavity-mirror's mode into higher order cavity modes. As in a classical feedback effect \cite{PI_general_approach} both optical modes act via radiation pressure on the cavity mirror. An instability can occur if two conditions are met: First, the frequency difference of the resonant higher order mode and the fundamental mode must be close to a mechanical eigenmode of the mirror. Second, the beat note of both optical mode shapes must have an important spatial overlap with the mechanical mode shape.
The risk to obtain this non-linear coupling between the optical energy stored in the Fabry-Perot cavity and the mechanical energy in the cavity mirrors increases by increasing the intra-cavity power.\\
In this work we present a high speed, high power 2D beam steering system based on two acousto-optic modulators (AOMs) as an important step in the development of an active mitigation strategy of PI in GW detectors.\\
Rapid beam steering is crucial in many advanced applications like optical coherence tomography \cite{oct}, holographic memory systems \cite{hologram}, optical switches and interconnects \cite{switches} and 3D laser imaging microscopy \cite{scanningMic}. 
Beam deflection can be achieved by a variety of different approaches which can be grouped in two categories. On the one hand there are mechanically driven deflectors like rotating polygon mirrors \cite{marshall,bass}, galvano-scanners \cite{marshall,bass} and piezo-scanners \cite{marshall,bass}. On the other hand there are deflectors that change the optical properties of an optically transparent medium with external forces or combine several beams to change the direction of the propagating beam in this medium. Examples are spatial light modulators \cite{SLM}, optical phased arrays (OPAs) \cite{phased-array}, electro-optic modulators \cite{roemer} and acousto-optic modulators (AOMs) \cite{AOM_deflection}. An overview of the different laser scan methods and their classification can be found in \cite{gorog,marshall,bass,roemer}. 
While mechanically driven deflectors often provide a large scan range they are limited in speed because of inertia associated with the mass of moving parts. 
High deflection speed in one dimension can be obtained for example with OPAs and was demonstrated for 40 MHz steering speed but for a small scan range of 4.2 mrad \cite{OPA12}. An OPA system has been developed with a larger scan range of 19 degrees (331 mrad) but with a lower speed of 2 MHz \cite{OPA20}.
Commercial electro-optic modulator based on KTN crystals have a steering speed of 100 kHz with a deflection range of up to 150 mrad \cite{KTN_deflector}. There has also been a one dimensional electro-optic system demonstrated with scanning speeds of up to 500 MHz at the expense of a reduced scan angle of 1-2 mrad \cite{KTN_16}. A 2D electro-optic deflector has been developed with a 13 MHz scan speed and a scan range of 2-3 mrad \cite{KTN17} which has been improved up to 7 mrad \cite{KTN19}.
Even though these systems offer a fast deflection they were tested for relatively low optical power ($\sim$ mw) and in sinusoidal scan mode.
A way to deflect high power beams in arbitrary directions are AOMs. 
AOMs deliver beam deflection by diffracting light: A transducer creates a propagating acoustic wave in the AOM's crystal inducing a spatial modulation of the refractive index which acts like a grating for the incoming optical beam. Depending on the application, the design of the crystal and of the transducer changes.
Commercial AOMs designed for deflection, i.e. AOM deflectors, have scan ranges up to 50 mrad with a steering speed of hundreds of kHz up to some MHz depending on the beam diameter.
AOM designed for amplitude modulation, i.e. AOM modulators, attain higher repetition rates up to hundreds of MHz at the expense of the scan range \cite{aa_optoelectronic}. 
Fast laser scanning with acousto-optic modulators has been investigated in the past for laser-based displays \cite{displayrev,korpel}, in microscopy\cite{micro2020} and imaging systems \cite{katona} and in beam shaping systems \cite{akemann,shape}.

\section{Active mitigation strategy of optomechanical parametric instabilities based on radiation pressure}
\label{sec:PI}

PI in GW detectors were predicted in 2001 \cite{Braginski,Braginski_2} and observed for the first time in the LIGO detector in 2015 \cite{PI_obs}.
Next generation ground-based GW detectors like the Cosmic Explorer \cite{Cosmic_explorer} and the Einstein-Telescope \cite{ET} intend to increase the sensitivity by a factor of 10 compared to current GW detectors. Since one key action to increase the sensitivity consists in increasing the optical power in the interferometer a larger number of PI are expected compared to second generation detectors.

Several PI mitigation strategies have been proposed \cite{PI_strategies}: There are, on one hand, passive strategies that aim to avoid PI by changing the cavity properties and, on the other hand, active strategies that monitor the onset of PI and 
mitigate it in a feedback loop.The first passive strategy that was used with success during the first observation run of 2nd generation GW detectors consists in changing the radius of curvature of the mirrors to avoid resonance of the higher order optical mode involved in PI \cite{thermal_tuning}. 
But in a higher-power regime, because of the high density of mechanical 
modes, PI involving other modes can still arise.
Another passive strategy that has been used during the last observation run is to increase the loss of mechanical modes that are candidates for PI by attaching little dampers to the cavity mirrors \cite{dampers,Biscans2019}. But implementing such dampers for a larger group of mechanical modes relevant to PI, would also increase the thermal noise level of the mirrors which limits the future generation of GW detector. An active technique which was tested in the LIGO interferometer consists in applying a damping force on the mirrors comprises damping of mechanical modes involved in PI by means of the electro-static drive actuators \cite{ESD,BlairPRL2017}, normally used to control the mirror position. But this technique is limited in the number of modes that can be efficiently damped because of the fixed position of the actuators.

\begin{figure}[htbp]
\centering\includegraphics[width=12cm]{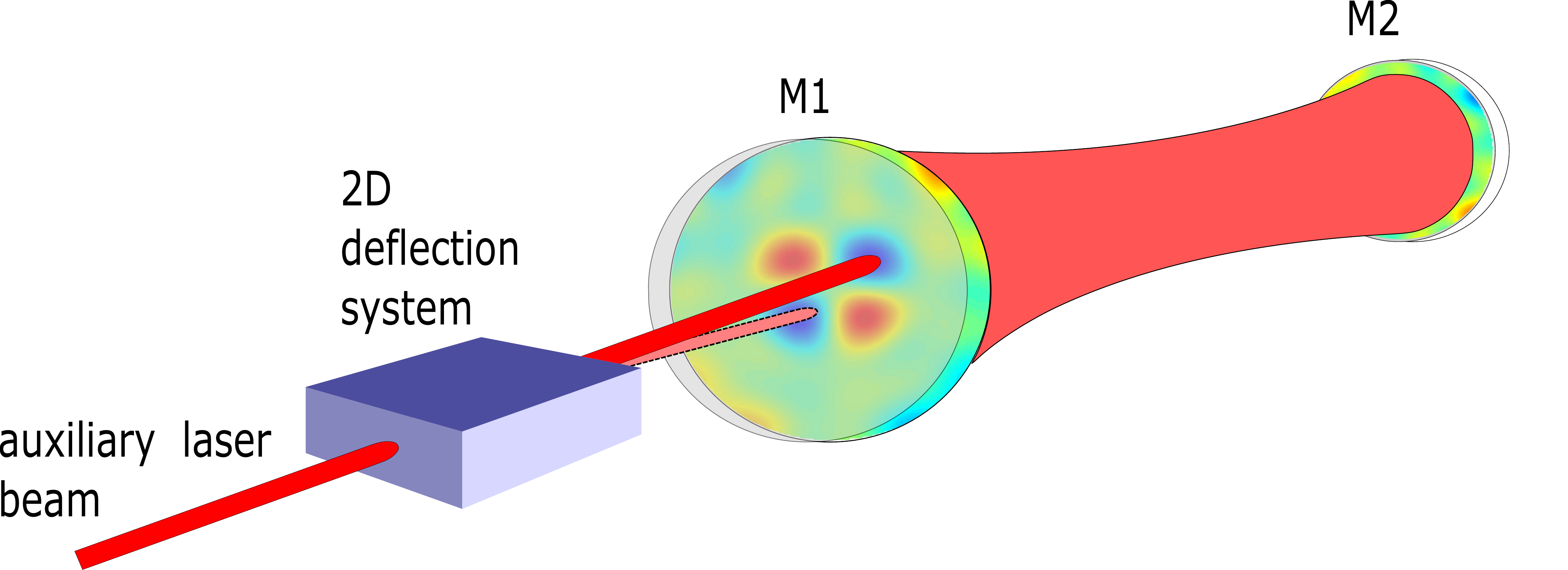}
\caption{Scheme of the PI mitigation strategy based on radiation pressure. Each arm of a gravitational wave detector interferometer contains a Fabry-Perot cavity consisting of two mirrors: M1 and an M2. Arising PI will be damped via radiation pressure by reflecting an auxiliary small beam at the high reflectively coating of mirror M1. The auxiliary beam will be controlled with the fast 2D deflection system to address the different lobes of the growing mechanical eigenmodes of the mirror.}
\label{fig:PI_mitigation_shema}
\end{figure}

We propose a flexible active PI mitigation strategy based on radiation pressure. Its principle is depicted in Fig.\ref{fig:PI_mitigation_shema} and a first presentation of concept can be found in \cite{PI_mitigation}.
This strategy consists in damping the rising mechanical mirror mode due to PI via a radiation pressure force applied by a small and movable auxiliary laser beam.
The small laser beam is deflected rapidly to address different lobes of the mechanical mode shape during one period. 
While pointing on one lobe the optical power will be modulated with the frequency of the mechanical oscillation with a phase shift of $90^\circ$ to damp the mirror displacement.

An estimation of the laser power $P$ needed to reduce the quality factor of a mirror oscillation which has an amplitude 10 times higher than in thermal equilibrium is presented in \cite{PI_strategies}:
\begin{equation}
     P=\frac{cF_0}{2}\frac{1}{B}=\frac{10c\sqrt{2m\omega^2\kappa_B T}}{2Q_f}\frac{1}{B},
 \end{equation}
  with the speed of light $c$, the required radiation pressure force $F_0$, the effective mass of the mirror which contributes to the mirror motion $m$, the angular frequency of the mechanical oscillation $\omega$, the Boltzmann's constant $\kappa_B$, the mirror temperature $T$ and the desired quality factor $Q_f$. The coefficient $B$ is added in this formula and represents the spatial overlap of the exerted radiation pressure force with the mechanical mode shape.
 Assuming values from current gravitational wave detectors \cite{PI_strategies,Blair} with $m=10$ kg, $\omega= 100$ kHz, $Q_f= 10^5$ and $T= 300$ K, the required auxiliary laser power is $P=0.43$ W which corresponds to a radiation pressure force of 2.9 nN for an overlap factor $B$ equal 1. Since the auxiliary laser beam has a small spot size and will scan the mechanical mode shape, the overlap coefficient $B$ does not only include the spatially averaged distribution of the applied force but it has also a temporal dimension and will be smaller than 1. The observed time constants of acoustic modes involved in PI where measured to be between 50 ms to several hundreds of seconds \cite{PI_strategies}, so if the rising mechanical modes of the mirror will be detected and damped in a feedback control before its amplitude rises ten times higher than thermal equilibrium, then even smaller forces should be sufficient.
Unlike in the technique based on electro-static actuators \cite{ESD,BlairPRL2017}, a large variety of modes can be damped because the force can be applied with a good overlap with any mode shape by the movable beam. Besides, more mechanical modes can be potentially damped at the same time.\\
Moreover, the mitigation beam applied to one of the cavity mirrors is able to damp any PI involving the other one. In fact, introducing a high order mode into to cavity was used to generate a PI \cite{BLAIR2013} or to damp it \cite{Zhao2015}. Instead of injecting the high order mode into the cavity, the latter can be generated by scattering the intracavity fundamental mode by reshaping the mirror surface using the mitigating beam. Properly reshaping the mirror surface can also scatter the high order mode involved in the PI into a non-resonant optical mode. Depending on the cavity configuration and the considered PI, one of these two optical mitigation methods can be used. \\


Summarizing, the proposed flexible active mitigation strategy demands a 2D beam deflection system with the following characteristics:
\begin{itemize}
    \item \textbf{high speed}: aiming at damping mechanical modes with frequencies ranging from tens to hundreds of kHz, the auxiliary laser beam should be deflected with a rate of some MHz to act on several lobes of the mode shape during one period of the highest mechanical mode frequency.
    \item \textbf{in random access}: in order to optimize the overall overlap of the exerted radiation pressure force and the mechanical mode shape, the beam deflection pattern should be easily adjustable.
    \item \textbf{high power}: the required damping force of some nN demands a deflected laser beam of the order of 1 Watt. 
    \item \textbf{large scan range with a small beam spot on the mirror}: a small ratio of the beam divergence to the angular scan range is important. The large scan range is necessary to access every point of the mirror. At the same time the beam spot should be small compared to the mirror diameter in order to address even small lobes of a mechanical mode shape with a high overlap.
\end{itemize}
In the following we will present the experimental investigations and the results obtained concerning these four points (Secs. \ref{sec:setup} and \ref{sec:results}). The possible layout scheme and the auxiliary laser requirements for the implementation on GW detectors will be discussed in Sec.\ref{sec:gwd}. 
\section{Experimental layout}\label{sec:setup}
\subsection{Optical setup}
\begin{figure}[htbp]
\centering\includegraphics[width=12cm]{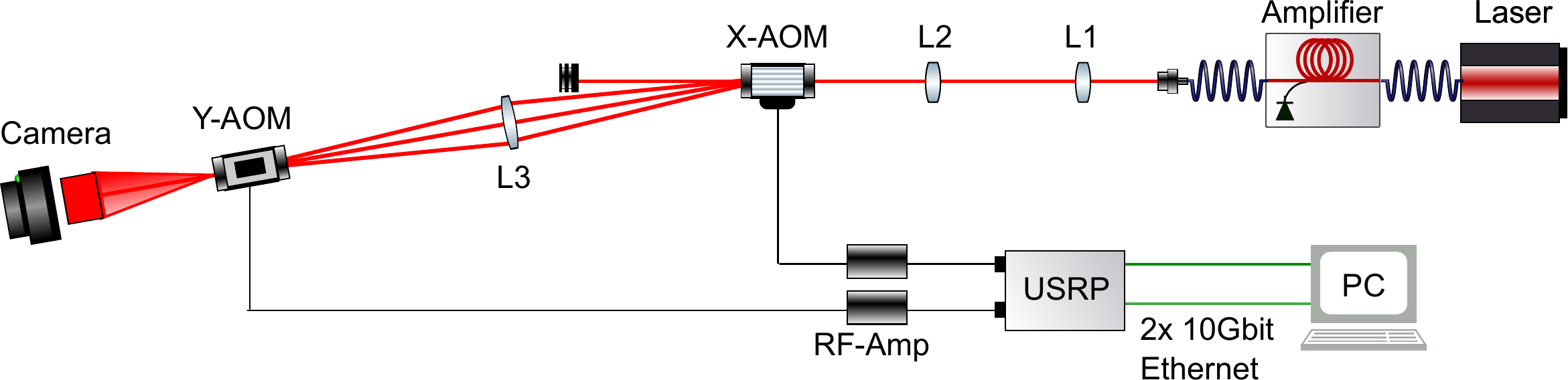}
\caption{Experimental setup. The seed laser is amplified and focused  with lenses L1 and L2 into the X-AOM. The zero-order diffracted beam is blocked with a beam dump. Lens L3 serves as imaging lens to enter the horizontal deflected beam spot into the Y-AOM. The 2D deflected beam is captured with a camera. The driver signal for both AOM is compiled with a PC, sent via dual 10 Gbit Ethernet connection to the USRP which creates the radio signals. These are amplified with RF amplifiers (RF-Amp) before reaching the AOMs. }
\label{fig:experimental_setup}
\end{figure}

In our experiment two AOMs (Gooch \& Hausego AODF-4224-2, 34 mrad deflection range, 130 MHz acoustic bandwidth) are lined up orthogonal to each other for fast 2D laser beam deflection (referred to as X-AOM and Y-AOM ). These AOMs are chosen as a compromise between amplitude modulation and high deflection angle: they fulfil both requirements of high speed (33 ns of rise time for 210 \textmu m beam waist) and a small ratio of the beam divergence (about 6 mrad) to the angular large scan range (34 mrad). Fig.\ref{fig:experimental_setup} shows the optical setup. We use a 10 mW seed laser (NKT Photonics Koheras Adjustik) with a wavelength of $\lambda$=1064 nm and a Azurlight Systems fiber amplifier (maximum output power: 50 W). With the set of two lenses (L1+L2) the beam is focused into the X-AOM to a diameter of 190 \textmu m. The casing of both AOM is relatively big and the active aperture of 300 \textmu m is too small to put one AOM directly behind the other one to ensure a deflection in two dimensions. For this reason an imaging lens L3 with a focal length $f=200$ mm is placed between the X- and
Y-AOM with a distance of 2f-2f to image the beam spot of the X-AOM with unity magnification into the Y-AOM. L3 is put into the optical axis of the first order diffracted beam at the central acoustic frequency of the AOM (225MHz). With a camera (CCD Camera Beam Profiler,BC106N-VIS/M, Ø30 \textmu m - 6.6 mm, 350 - 1100 nm) the diffracted beam profile is captured and analyzed.

The following considerations for the choice of the imaging lens L3 were taken into account: the distance between the final Gaussian beam waist position and the geometrical conjugate plan increases by reducing the focal length of the chosen lens\bibnote{The relative deviation between the geometrical image of the object waist and the final waist position is proportional to $1/f^{2}$}. If this difference is close or larger than the Rayleigh length of the Gaussian beam it will affect the deflection efficiency of the Y-AOM since the incoming beam does not have a flat wave front. But, on the other hand, a large focal length demands a large lens diameter in order to collect the entirety of the deflected beam at a distance of 2f after the X-AOM and it would increase the size of the whole system. The calculation of Gaussian beam propagation shows that, for a lens with focal length of $f=200$ mm, the final waist lies 4 mm before the geometrical conjugate plan which is acceptable as it is within the Rayleigh length of 26 mm for a spot waist diameter of 190 \textmu m. A lens diameter of 2 inches is large enough to collect the beam over the deflection range. The Gaussian beam profile is preserved after deflection (see Fig.\ref{fig:36_deflection_spots}) meaning that aberrations of the imaging lens are negligible.
Another possible configuration for unity magnification imaging of the beam spot consists in using a set of two lenses with the same focal length in a 4f imaging system \cite{scott}. But practically it is not evident to find two lenses with the exact same focal length which would result in a spot magnification unequal to one.

\subsection{Generation of the driver signals of the AOMs}
\label{generation_RF}
The creation of fast changing radio frequencies (RF) signals in random access is a challenge. Commercial AOM drivers are either voltage controlled oscillators (VCO) with rather slow commutation times in the \textmu s range \cite{aa_optoelectronic} or direct digital synthesizers (DDS) with commutation times of hundreds of ns \cite{aa_optoelectronic,isomet,gooch_housgo}. Additional delay is added by programming data to the DDS chip and is limiting the maximal output rate. A frequency generator based on a field-programmable gate array (FPGA) and a DDS has been presented with consecutive frequency change of less than 1 \textmu s \cite{FPGA_DDS_2017} even if the frequency change at the DDS is done below 100 ns \cite{FPGA_DDS_2015}. These solutions are not fast enough for our application in order to be only limited by the nominal rise time of the AOM which is of 33 ns for a 210 \textmu m beam spot.

We used an Universal Software Radio Peripheral (USRP X310,KINTEX7-410T FPGA, 2 channels, 160 MHz bandwidth each channel, dual 10 Gigabit ethernet, Ettus Research) a software-defined radio for the radio frequency generation. RF signals generation is based on In-phase (I) and Quadrature (Q) modulation (with a bandwidth of 160 MHz) of a carrier signal ranging from DC to 6 GHz.
The hardware architecture consists of a large FPGA with two daughterboard slots including digital analog converters that generate the modulation signals and of an analog frequency generator for the carrier. The USRP has a nominal sample rate of 200 MSamples/s which allows to change the transmitted frequency theoretically within 5 ns.
But it is not evident to attain this performance. Special care must be taken in optimizing the computer interface and the FPGA configuration.

The USRP is interfaced with the USRP Hardware Driver\texttrademark\,
(UHD) software on a linux machine (Dell Precision 5820). A dual 10 Gigabit ethernet connection between the USRP and the host is established to enable a 2x 200 Msamples/s transmission with 32 bits per samples ( total data transfer of 1.6 GB/s). To enable the highest streaming rates over a network connection a list of configurations needed to be done on the host: The CPU governor is set to performance and priority scheduling is enabled to use the whole CPU capacity for the data transmission. The network buffers are set to a maximal value of 625,000,000 bits and jumbo frames of 8000 bytes are set as maximal transmission unit (MTU) for the network transaction layer to maximize the data throughput. For further acceleration a data transport based on data plane development kit (DPDK), which uses user space memory buffers for the data communication over an network interface, is set up.


On the USRP side the shipped FPGA image has been modified by adding two additional First In First Out (FIFO) buffers to enlarge the bandwidth of the already existing direct memory access FIFO buffer and DPDK transmission was enabled.\\
In this configuration a set of frequencies and their duration is loaded into UHD and sent continuously in a loop to the USRP. Two RF amplifiers (MPA-40-40, 20-1000 MHz, 4 W, RF Bay Inc.) are used to increase the RF level to the maximal AOM input power of 3 W.\\ 

\begin{figure}[htbp]
\centering\includegraphics[width=1\linewidth]{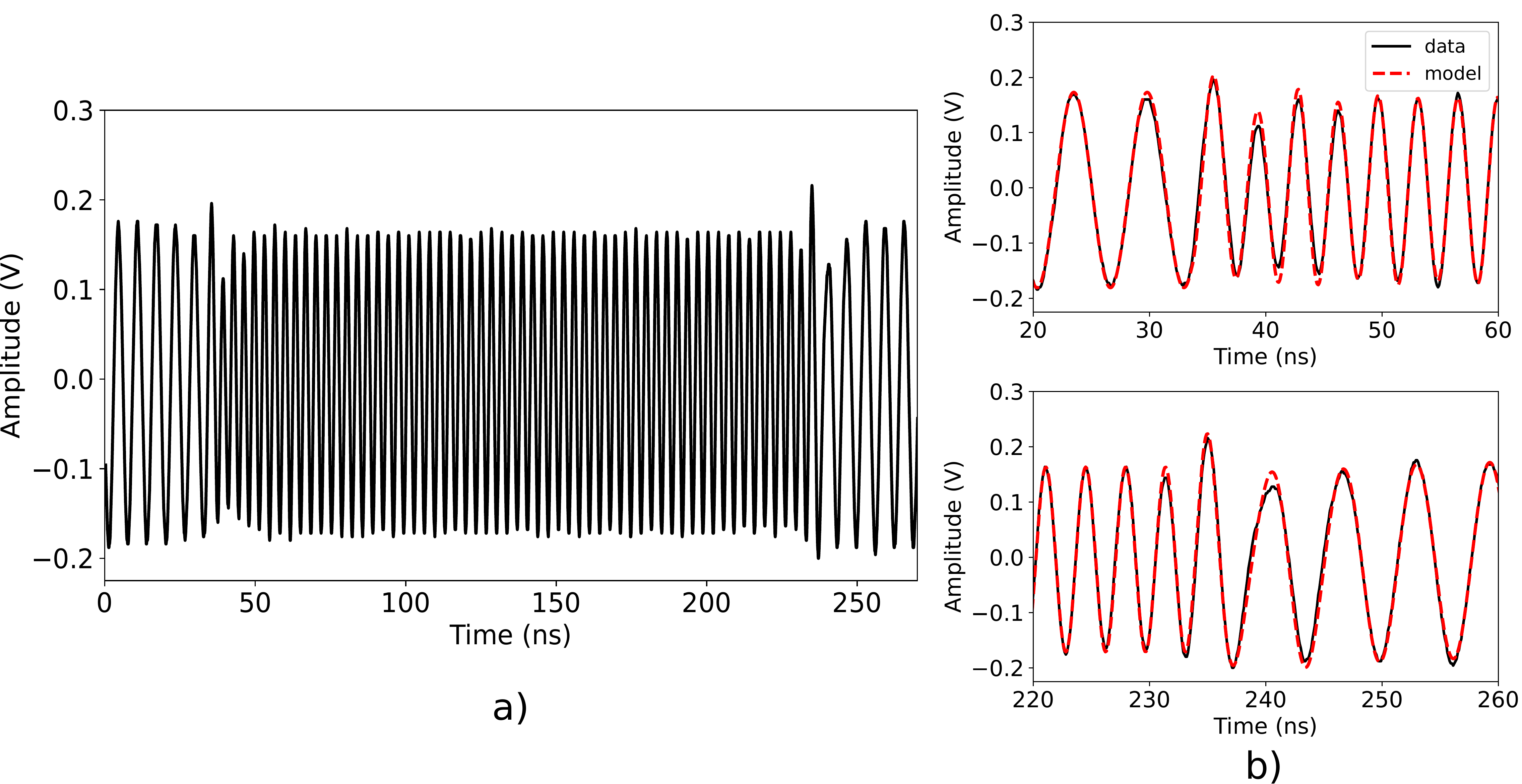}
\caption{a) Portion of the signal transmitted by the USRP. The frequency changes every 200 ns between 160 and 290 MHz. b) Zoom on the frequency transitions: 160 to 290 MHz (top), 290 to 160 MHz (bottom). An exponential function is used to model the frequency switch, the fit is plotted in red dashed lines.}
\label{fig:USRP_transition_160_290MHz}
\end{figure}

A sample of the USRP's transmitted signal with a frequency change between 160 MHz and 290 MHz every 200 ns is shown
in Fig.\ref{fig:USRP_transition_160_290MHz}(a). Two frequency transitions are visible and are zoomed-in in
Figs.\ref{fig:USRP_transition_160_290MHz}(b). One can see that when the signal frequency changes, the signal amplitude fluctuates. We defined a simple mathematical model to characterize these transitions. The goal is to estimate the time required to change the frequency and the time required to reach a stable amplitude. The fit of the model to the experimental data is shown in dashed lines in Figs. \ref{fig:USRP_transition_160_290MHz}(b).
The model and the detailed analysis can be found in \textcolor{blue}{Supplement 1}. 
According to our model, the signal frequency takes a characteristic exponential time of about 2 ns to change for both transitions while the characteristic time for the signal amplitude stabilization is about 7 ns . 
This analysis confirms that the USRP fulfils the requirements on the AOMs driving signal: it provides RF signals in random access with a transition time much lower than the AOM rise time. The speed of our 2D deflection device will be then limited by the AOMs' response as we'll see in the next section.

\section{Results}\label{sec:results}

\subsection{Deflection efficiency}
\label{sec:DE}
\begin{figure}[h]
    \centering
    \subfigure[]{\includegraphics[width=0.49\textwidth]{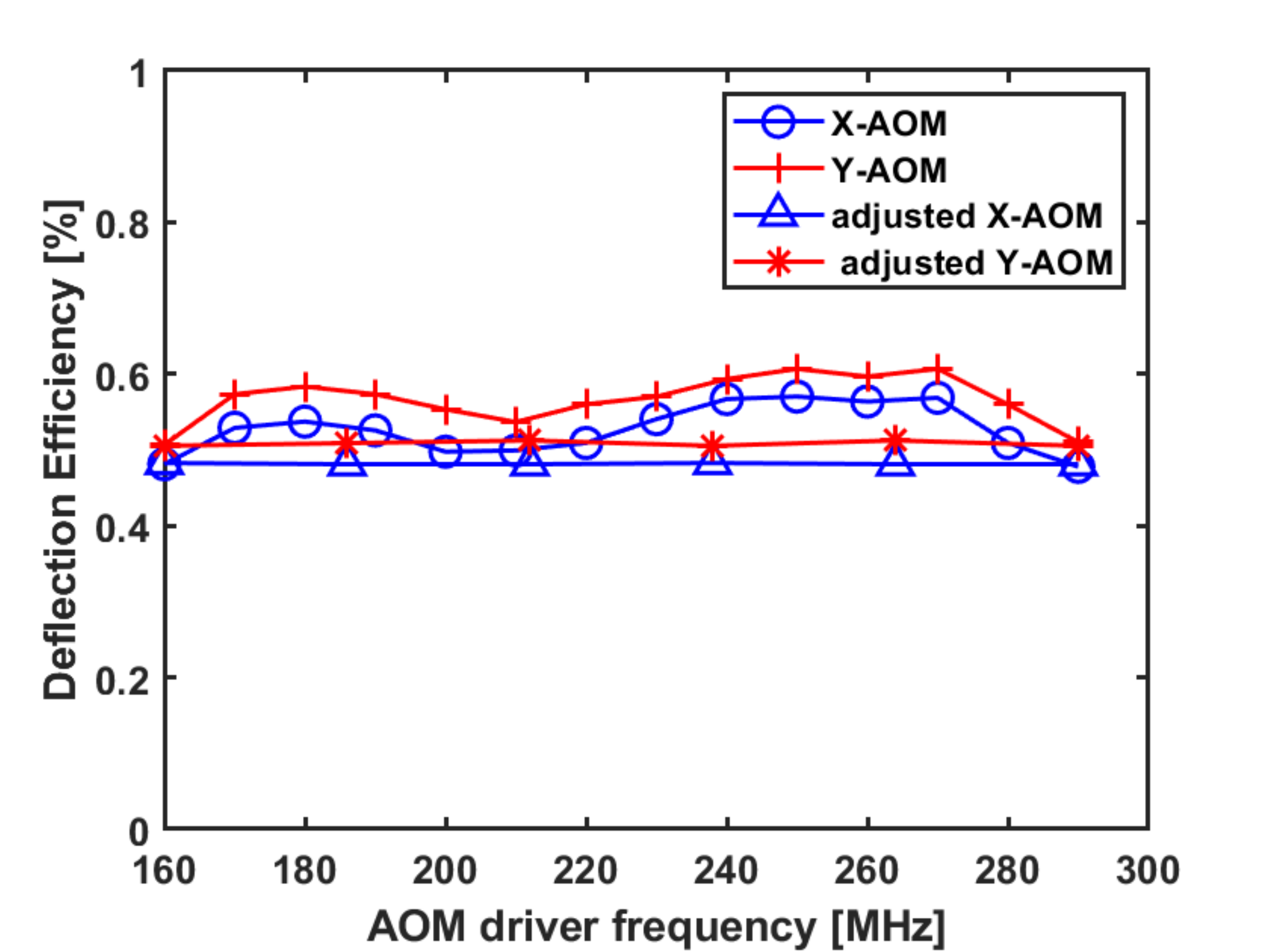}\label{fig:DE_for_both_AOM}} 
    \subfigure[]{\includegraphics[width=0.49\textwidth]{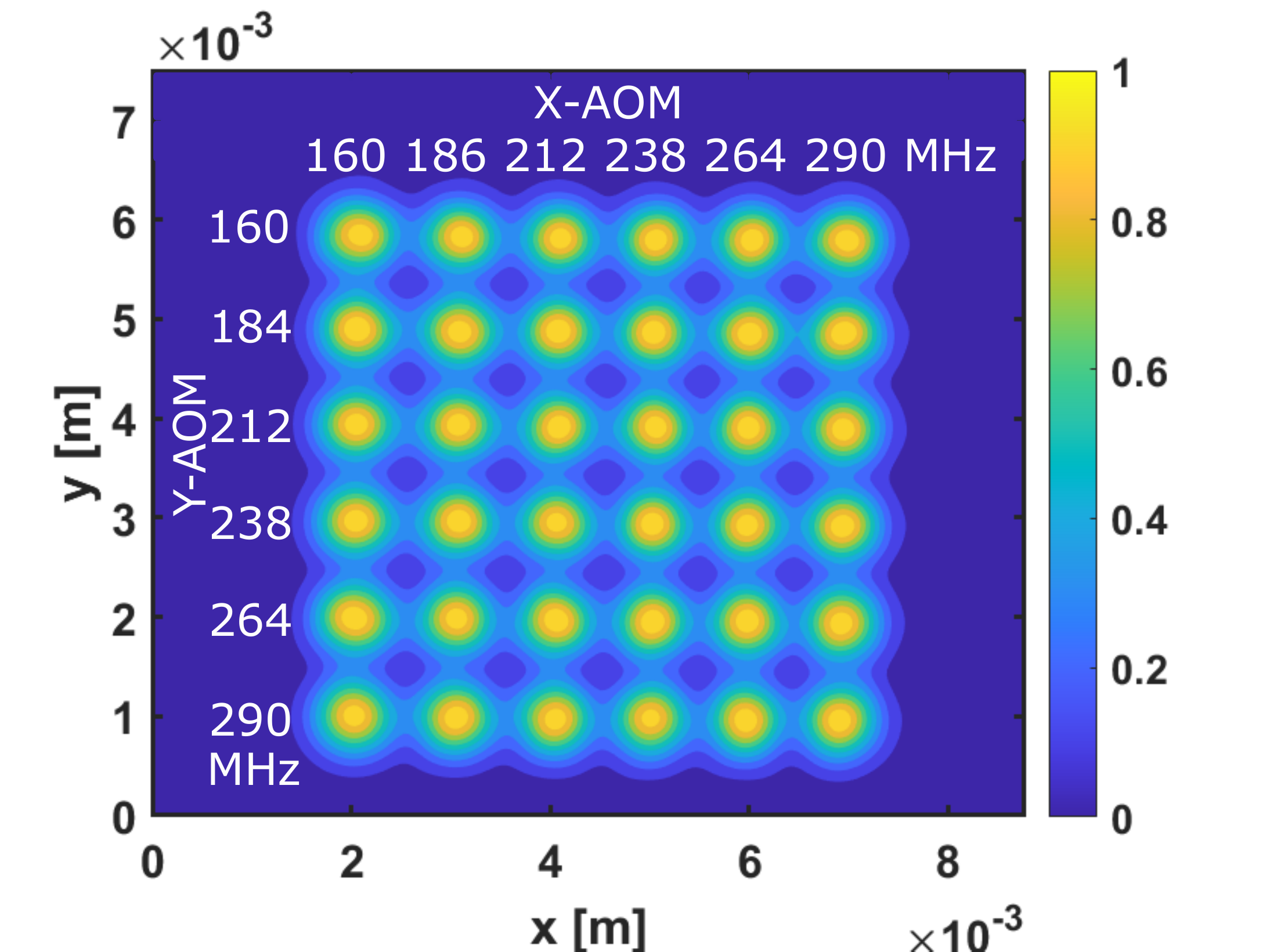}\label{fig:36_deflection_spots}} 
    \caption{Deflection performances of both AOMs .a) DE curves optimized for the highest and flattest values in the whole acoustic bandwidth of the AOMs. Before (blue circles and red crosses) and after adjusting the RF power to obtain the flattest curve (blue triangles and red stars).
    b) 2D deflection scan pattern captured with a camera 15 cm after Y-AOM with an exposure time of 836 ms. The beam is moved every 3 \textmu s, column by column, from the top-left position. The corresponding driver frequencies for both AOMs are shown. The optical power is normalized.}
    \label{fig:deflection_efficiency}
\end{figure}

Both AOM are mounted on a multi-axis stage to optimize the tip tilt and rotation angle for a flat deflection efficiency (DE) curve in the whole acoustic bandwidth of the modulator from 160 to 290 MHz.  Fig.\ref{fig:DE_for_both_AOM} shows the measured efficiency curve for the X- (blue circles) and Y-AOM (red crosses) for a beam spot diameter of 190 \textmu m and a RF power of 2.7$\pm$0.4 W. The mean deflection efficiency of the X- and Y-AOM is 52.7\% and 56.6\% with a standard deviation of 6\% for both AOMs. The difference in deflection efficiency between the two AOMs is considered to be due to manufacture differences since we tested both AOMs under the same conditions. Higher deflection efficiencies can be obtained with these AOMs but for a smaller driver frequency range. But, as we aim at having the largest possible scan angle with a flat deflection efficiency curve, we chose this working point.
To obtain even a flatter deflection efficiency through out the acoustic bandwidth of the AOM the RF power of each driver frequency was adjusted in the UHD software to get the same deflection efficiency as the driver frequency with the lowest efficiency. Based on the deflected optical power measured with a power meter the needed RF power for each driver frequency is calculated, set in the software and adjusted by measuring the new deflected beam power again. The flat efficiency curves obtained for X- (blue triangles) and Y-AOM (red stars) are also displayed in Fig.\ref{fig:DE_for_both_AOM}. The mean values are respectively 47.5\% and 51\%.\\
For the 2D deflection, a set of driver frequencies for both AOMs is chosen in a way to include as many deflection spots as possible in the whole scan range. Considering that the deflected beam will be used far beyond the Rayleigh length, the ratio of the scan angle (nominally 34 mrad) to the beam divergence gives the number of spots in 1 dimension. The closest integer to this ratio is 6 so the 2D deflection pattern has 36 spots. The distance between two consecutive beam spots corresponds to the beam diameter (at around 13\% of the peaks' amplitude).\\
The resulting scan pattern is captured with a beam profile camera 15$\pm$1 cm after the Y-AOM with an exposure time of 836 ms and is presented in Fig.\ref{fig:36_deflection_spots}. 
The scan range of both AOMs in this plot is around 4.9\textmu m (peak to peak) which equals to an angle of 32$\pm$2 mrad. This is consistent with the deflection angle defined by the manufacturer.\\
The RF power adjustments allow to obtain an homogeneous power spatial distribution.
We measured the optical power of each spot position and obtained a standard deviation of less than 1\% of the mean spot power. The DE of the whole 2D deflection system is 24.2\% . Moreover each spot shows a well conserved transverse Gaussian beam profile which indicates that both AOM and the imaging configuration don't introduce any beam distortion.

\subsection{Deflection transition time}

\begin{figure}[htbp]
\centering
    \subfigure[]{\includegraphics[width=0.49\textwidth]{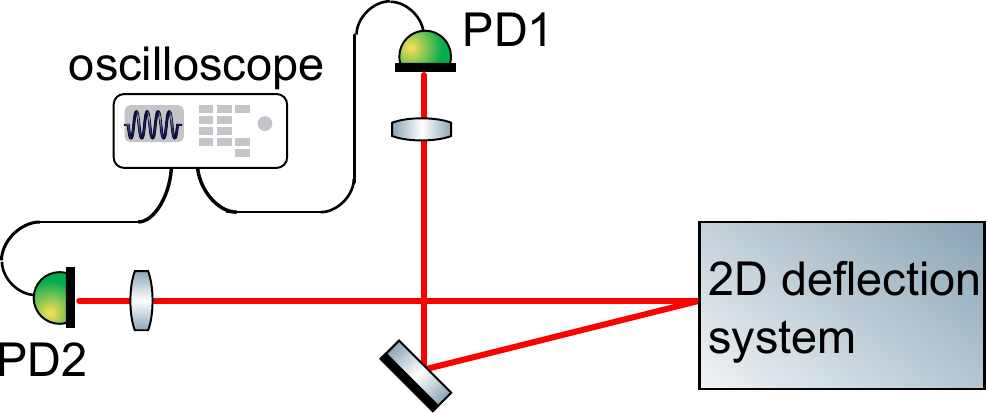}\label{fig:transition_time_setup}} 
    \subfigure[]{\includegraphics[width=0.49\textwidth]{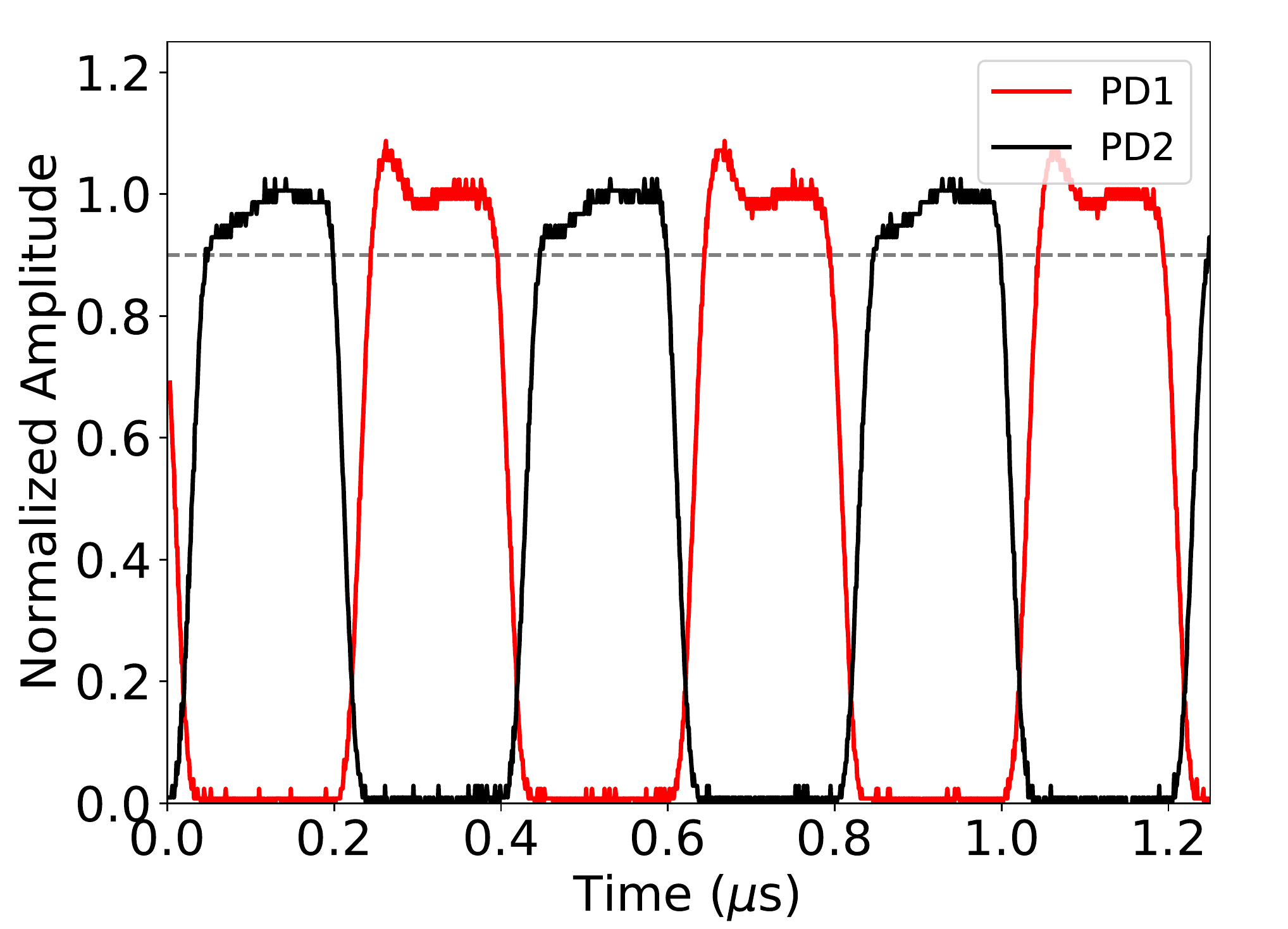}\label{fig:transition_time_200ns}}
    \subfigure[]{\includegraphics[width=0.52\textwidth]{risetime2}\label{fig:riset}}
\caption{a) Experimental setup to measure the transition time for beam deflection in random access. Two photodiodes (PD1 and PD2) are used to detect the deflected beam on the positions corresponding to a driving frequency in both AOMs of 160 MHz (PD1) and 290 MHz (PD2). The AOMs' driving frequency is switched continuously every 200 ns. The AOMs input spot diameter is 190 \textmu m. b) Normalized signals of PD1 (red) and PD2 (black). The dashed line indicates the 90\% of the stationary signal level used to measure the transition time. c) Zoom of (b) with fit of a second order system step response (blue straight line)}
\label{fig:transition_time}
\end{figure}

The transition time of consecutively deflected beams gives the rapidity of the system and was experimentally measured. For this measurement we chose two random deflection directions. Two fast photodiodes with a bandwidth of 200 MHz were placed, as illustrated in Fig.\ref{fig:transition_time_setup}, after the 2D deflection system. For this measurement the laser beam was continuously deflected between the spot on the top left and the bottom right in Fig.\ref{fig:36_deflection_spots} staying on each position for 200 ns.
For each photodiode a lens is used to focus the entire beam on the sensor. The corresponding driver frequencies for (X-AOM,Y-AOM) are (160, 160) MHz and (290, 290) MHz. 
The driver frequencies for both AOMs are synchronized and create acoustic waves in the AOM which propagate with 4200 m/s through the Tellurium Dioxide Crystal. In order to synchronize the deflection of both AOMs the beam needs to hit the crystal of both AOMs
at the same distance from the transducer, neglecting the light propagation time. This can be achieved by moving one AOM parallel to the axis of the acoustic wave propagation in its crystal with a translation stage. In this way the shortest transition time between two randomly chosen deflection positions is obtained. After this optimization, we expect that the dynamics is limited by the AOMs response because the USRP's response have shorter time constants (see Sec.\ref{generation_RF} and \textcolor{blue}{Supplement 1}).
Figs.\ref{fig:transition_time_200ns} and (c) show the signals measured by both photodiodes. When the beam is on photodiode 1 (PD1), the signal of PD2 equals 0 and vice versa. One can also see that the deflected beam power takes some time to reach the final level and in a different way for PD1 and PD2. 
The temporal evolution of the deflected beam power on PD1 position is described quite well by the step response of a second order system. The same description fits also rather well the signal of PD2 but with some discrepancies with the data and with a chi-square twice as big (see fit in Fig.\ref{fig:riset} and \textcolor{blue}{Supplement 1} for details). Based on the fit, the rise time is measured respectively 39 ns and 45 ns for PD1 and PD2. To evaluate the transition time we measure the time interval from when the signal of one PD reaches its 90\% level during the fall to when it reaches 90\% of its level on the other PD. The measured transition time is 50$\pm$5 ns. This means that the beams stays effectively around 150 ns at the addressed position with an intensity of more the 90\%. By reducing the duration of each driving frequency to 100 ns it gives an effective dwelling time on a given position of about 50 ns (amplitude higher than 90\% of maximum value). This is the minimum dwelling time that can be obtained without reducing the beam power thus the applied force. The maximum deflection rate is then 10 MHz. An additional Pockels cell in the experimental setup could be used to cancel the amplitude modulation observed above the 90\%.\\

\subsection{High optical power deflection}

\begin{figure}[h]
\centering
    \subfigure[]{\includegraphics[width=0.49\textwidth]{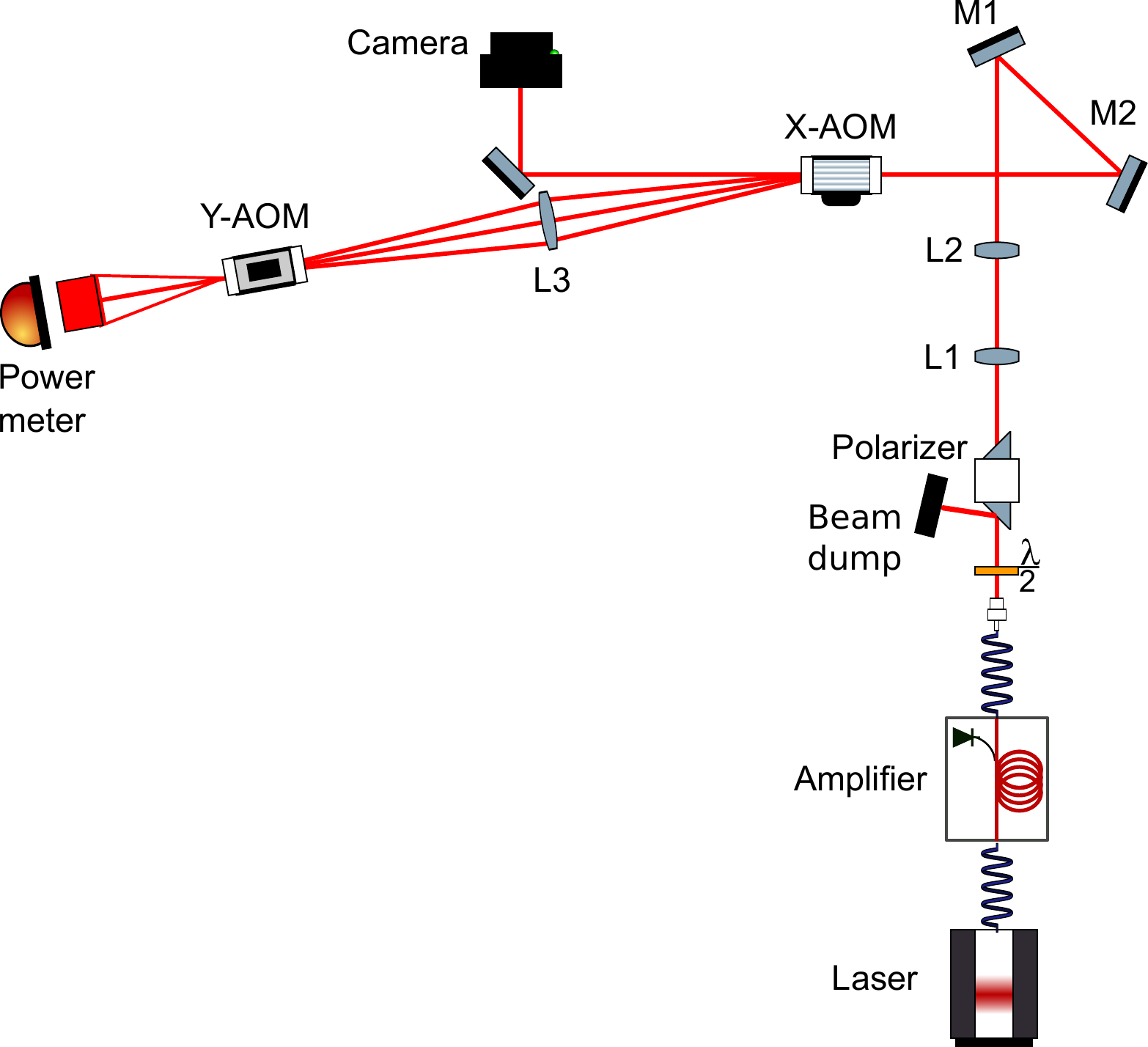}\label{fig:high_power_setup}} 
    \subfigure[]{\includegraphics[width=0.49\textwidth]{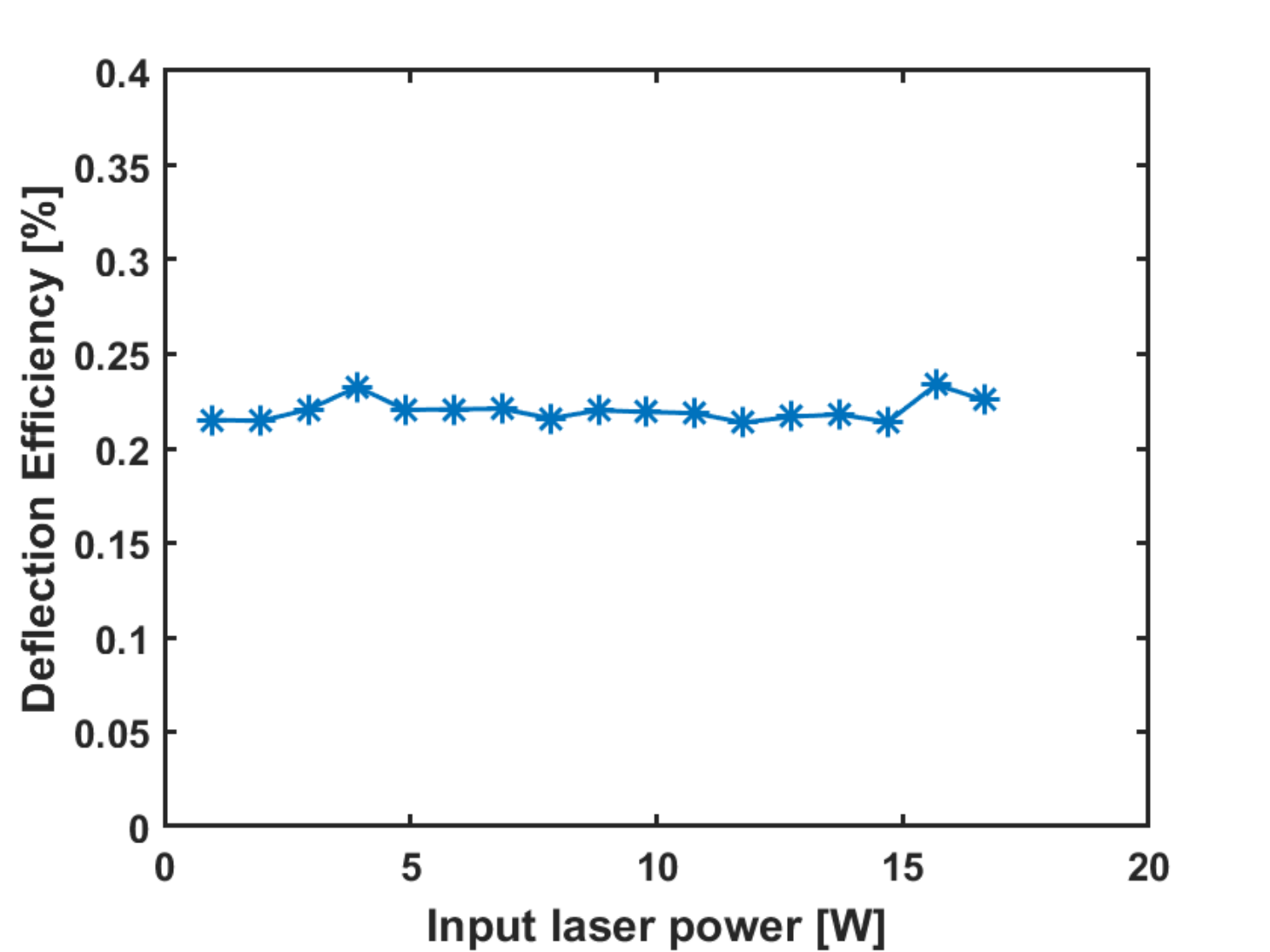}\label{fig:DE_high_power}} \caption{(a) Experimental setup used for high power beam deflection. The seed laser gets amplified with a 50 W fiber amplifier. A set of a half-wave plate ($\frac{\lambda} {2}$) and a polarizer are used to regulate the laser power. Lenses L1 (f=750 mm) and L2 (f=200mm) are used to focus the beam to a spot diameter of 200 $\mu m$ at the X-AOM. 
    Lens L3 (f=200 mm) images the spot of the X-AOM into the Y-AOM. The 0 order of the X-AOM is imaged with a camera. The final deflection power is measured with a power meter. (b) Total DE measured at different optical input power.}
\label{fig:high_power}
\end{figure}

High power beam deflection is important for our application of a laser beam as source of radiation pressure.
We used a 50 W fiber amplifier (Azurlight Systems) in order to determine the limits of our 2D deflection system. Fig.\ref{fig:high_power_setup} shows the optical setup of this experiment.
The optical power is adjusted by using a set of a half-wave plate and a polarizer. The beam spot is focused with a set of two lenses on 200 \textmu m into the X-AOM and the optical laser power is increased step by step.  With the manufacturer specification for the optical power density of 500 W$/\text{mm}^2$, the maximal laser power should be 15.7 W. A camera monitors the transmission beam shape of the X-AOM.
For the deflection efficiency measurement the AOM input power is measured between L1 and L2 and the power of the deflected beam of Y-AOM. The insertion losses of the optical components in between were determined and taken into account. Fig.\ref{fig:DE_high_power} shows the deflection efficiency of the whole 2D deflection system for different optical input power. We increased the laser input power to 16.7 W without facing any decrease in deflection efficiency. But we observed a thermal lens by increasing the input laser power with a spot size variation of 8\%. This measurement shows a flat deflection efficiency of 22\% for input powers up to 16.7 W for our 2D deflection system. The  maximal deflected beam power was 3.6 W. The fact that the efficiency is 2\% lower than the value measured in section \ref{sec:DE} can be due to laser power fluctuations and the fact that we used two different power meter for this power range with a nominal accuracy of $\pm 1\%$ defined by the manufacturer. 
\section{Perspectives towards implementation on a GW detector}\label{sec:gwd}
\begin{figure}[htbp]
\centering\includegraphics[width=0.6\linewidth]{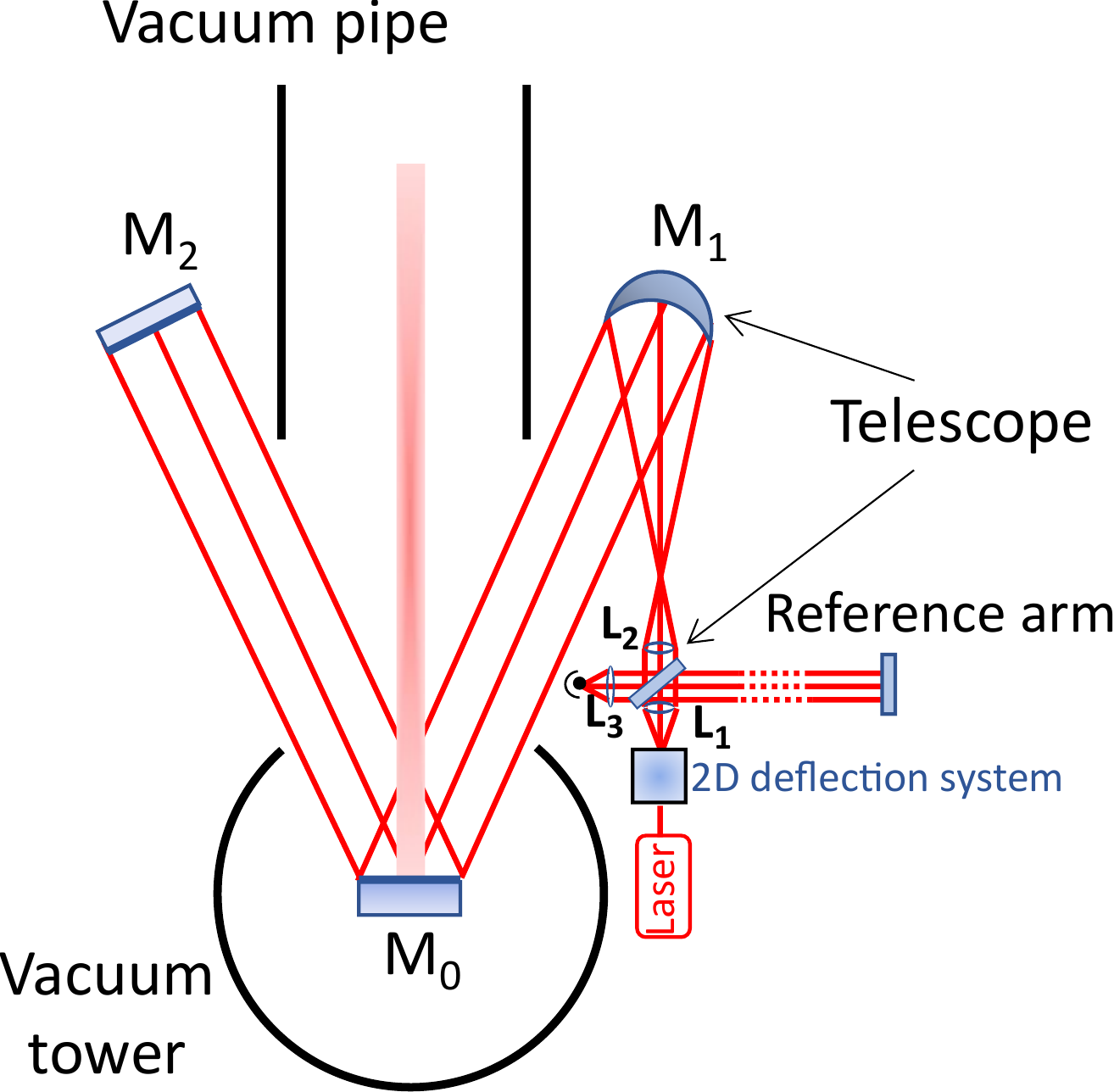}
\caption{M$_0$ is the suspended test mass; lens L$_1$ collimates the steering beam, L$_2$ and M$_1$ constitutes the telescope which adapts the collimated beam to the required range of the M$_0$; M$_2$ is the retro-reflection mirror}
\label{fig:dessin-joli}
\end{figure}
In Fig.\ref{fig:PI_mitigation_shema}, a simple scheme to illustrate the mitigation strategy was shown but in order to infer on the possible application in GW detectors we need to consider the available space around the suspended mirrors and the noise requirements on the auxiliary laser.\\ In transmission of the end mirror, for example, there are optics and photodiodes that are used for the interferometer control and that prevent an auxiliary beam to impinge perpendicularly from the rear surface. Therefore the 
implementation of the radiation pressure-based PIs mitigation system still requires an accurate feasibility study and might need a redesign of both the optical and the vacuum system around the test masses. In Fig. \ref{fig:dessin-joli} we propose a possible implementation which can be adapted to the currently used architecture of GW detectors and both for input and end mirrors. Its main feature is based on :
\begin{itemize}
    \item collimating the steering beam using the lens L$_1$ which allows to have possibly a bundle of parallel beams.
    \item the bundle goes through a Michelson interferomter used to sense the surface distortion of the test mass M$_0$. Ideally, this sensing Michelson works in a dark fringe configuration but arms are not required to be of the same length if the laser is phase locked to the interferometer main laser. At the output of sensing Michelson, the bundle is focused on the photodiode thanks to the lens L$_3$.  
    \item the reference arm in the sensing Michelson is used to lock the dark fringe at low frequency below the test mass resonances' frequencies. The second arm accommodates a telescope (lens L$_2$ and mirror M$_1$) which adapt the bundle to the test mass size. Since the bundle impinges on the test mass with an angle of roughly 30$^\circ$, the telescope is designed to shape all beams so it applies an axisymmetric radiation pressure on the test mass (L$_2$ can be slightly cylindrical). Steering angles need to be adapted.
    \item the bundle is retro-reflected by the mirror M$_2$ back to the test mass which allows to double the radiation pressure and then the efficiency of the mitigation process. The sensing Michelson is hence closed. The distance between the M$_0$ and M$_2$ has to be kept below few meters so the time propagation of light back and forth does not limit the beam steering rapidity. 
\end{itemize}

As for the electrostatic actuators, the radiation pressure is a one direction force and acts on the external degrees of freedom of the test mass. The applied force has both DC and AC components (at the mechanical mode frequency and at the steering frequency) ; precautions are then to be taken for both of them :
\begin{itemize}
    \item During the mitigation process, the DC component changes over time. In order to keep the cavity resonant, the DC component is compensated by the longitudinal control system.
    \item The DC component suffers from the residual force applied through the power fluctuation of the steering beam. Power fluctuations are characterized by  Relative Power Noise spectrum $\textrm{RPN}\left(f\right) = \delta P/P$. The requirements in terms of intensity noise of the PI damping beam should be similar to the ones for the Photon calibrator beam\cite{Pcal_ligo,Pcal_virgo}. By setting the acceptable relative displacement noise of the mirror at 10 times lower than the detector sensitivity, we can estimate the Relative Power Noise requirement $\textrm{RPN}\left(f\right) < \frac{c M\pi^2 h\left(f\right)L f^2}{10 P}$\bibnote{The formula is obtained from the displacement noise caused by the force fluctuations: $\delta x = \frac{\delta F}{M\left(2\pi f\right)^2}< h\times L/10$}. For the foreseen test masses in postO5 scenarios (prospective document not released yet) $M$ =105 Kg, the arm length $L$ =3 Km and the power of the steering beam is P =2 W. The most strict condition is dictated at 10 Hz to be $\textrm{RPN} < 1.4\times 10^{-7}\,\textrm{Hz}^{-1/2}$ for a sensitivity of $h\left(10\,\textrm{Hz}\right) = 3\times10^{-23}\,\textrm{Hz}^{-1/2}$, the maximum sensitivity being close to $h = 10^{-24}\,\textrm{Hz}^{-1/2}$. Although requiring a specific stabilizing system,  this stability is achievable specially compared to the power stabilization of the main laser at the level of few $10^{-9}\,\textrm{Hz}^{-1/2}$ for current Virgo and LIGO.
    \item The steering beam hits the mirror with an angle and transfers a transverse momentum to the test mass. Thanks to the mirror M$_2$, the retro-reflected transfers an opposite momentum which ideally cancels the first one. An imbalance between the two beams causes the DC transverse motion depending on the optical power increase during the damping process. Test mass's control can then be used to compensate this effect.
    Power imbalance or asymmetry on the impinging areas between the bundle beams is likely to provoke tilt and yaw motion of the test mass. This can be reduced by properly calibrating the beam power through the 2D deflection system and also the corresponding RF frequencies, for example by monitoring the bundle right behind M$_2$. Here also the test mass's control system can then be used to compensate this motion.
    \item All AC induced motions related to power imbalance or asymmetries happens at high enough frequency (outside the detection band) to not be a problem . However, one has to avoid exciting angular or translational resonance mode the frequency of which would match in this case the mirrors internal mode frequency. This would prevent the interferometer control.
\end{itemize}
In practice, the Michelson continuously monitor the status of the test mass mechanical mode at low power (10 mW). Once the amplitude exceeds the thermal level due to a PI, the power is increased to damp it while monitoring its amplitude. For a mechanical $Q$ factor $Q=10^7$, a thermal peak at 10 kHz will have an amplitude of $2.5\times 10^{-15} \text{m}/\sqrt{\text{Hz}}$. The sensing Michelson, will be limited by :(i) the shot noise at 10mW at $5\times 10^{-16} \text{m}/\sqrt{\text{Hz}}$(ii) thermal noises of the beam splitter and the reference arm mirror, at $10^{-16}\text{m}/\sqrt{\text{Hz}}$ corresponding mechanical resonances must be at different frequencies.(iii) Seismic noise and coupling of the frequency and RPN noises can be made negligible. 
The same sensing Michelson can be used to measure the quality factor of a single mode of the test mass following a ring down procedure and the corresponding spatial vibration amplitude on the test mass surface.
Finally, the wavelength of the auxiliary laser should not be resonant in the cavity but, at the same time, it should keep a reflectivity close to 1 on the test mass regarding the incident angle.
\section{Conclusion}
We demonstrate a rapid 2D beam deflection system which has all the characteristics required for active PI mitigation in GW detectors.\\
The fast deflection is based on AOMs which combine the large deflection range of an acousto-optic deflector with the rapidity of an acousto-optic modulator. The measured transition time between two random steering positions is 50$\pm$5 ns and it does not depend on the distance between the two positions. This allows a deflection rate of up to 10 MHz which will be necessary to damp mechanical modes with hundreds of kHz.\\
The random access deflection is enabled by a software defined radio device (USRP) that is used as fast AOM driver. The fact that the beam can be deflected in random access allows the implementation of an optimal deflection pattern to damp a large range of mechanical modes.\\
A deflection for a maximal beam power of 3.6 W is demonstrated which equals to a radiation pressure force of 24 nN. A first estimation shows that this force is sufficient to damp a rising unstable mechanical mode. 
These performances were obtained for a laser with a wavelength of 1064 nm and a spot diameter of 190 \textmu m in the AOMs. By decreasing the spot diameter in the AOM even faster transition times could be achieved with the cost of lower maximal power in the deflected beam due to power density limits of the AOM crystal. In this way the available radiation pressure force for PI mitigation would be decreased. In addition, a further focused beam will have a larger divergence which would mean the deflected beam hits the cavity mirrors with a larger spot diameter. \\
As mentioned in section \ref{sec:PI},
in order to address mechanical mode shapes with small lobes a small laser spot is desirable to obtain a large overall overlap. The important quantity that has to be small is the ratio of the beam divergence to the scan angle. We reached stable deflection performances over the whole scan angle of the AOM i.e. 34 mrad. Whereas the beam divergence results from the choice of the optimal spot diameter in the AOM that is a compromise between the rapidity and the maximal available radiation pressure force. The obtained ratio is 1/6 that means that the spot diameter on the target mirror surface is one sixth of the beam diameter.\\ 
The proposed PI mitigation approach is a promising, flexible and active strategy suitable for next generation GW ground-based detectors. They will inevitably face PI as they foresee to function with even higher optical power compared to current detectors. We also proposed a possible implementation scheme which uses the 2D beam steering system for PI mitigation and detection using a sensing Michelson. 
Further experimental investigations and simulations need to be done to better identify and quantify the impact of locally applied pressure on a suspended cavity mirror for PI mitigation.

\section*{Funding}
Agence Nationale de la Recherche(ANR-10-LABX-48-01, Labex FIRST-TF); Université Côte d'Azur (CSI-2019-COM4SPIN); Observatoire de la Côte d'Azur (BQR-2019-COM4SPIN).
\section*{Disclosures} The authors declare no conflicts of interest.
\section*{Data availability} Data underlying the results presented in this paper are not publicly available at this time but may be obtained from the authors upon reasonable request.
\section*{Acknowledgments}
The authors wish to acknowlegde Christophe Alexandre, Conservatoire National des Arts et Métiers (CNAM), Paris, for his advices on how to setup the PC and the USRP to obtain the highest performances. Special thanks goes also to Gilles Bogaert from our laboratory who contributed to the PI simulations.

\section*{Supplemental document}
See \textcolor{blue}{Supplement 1} for supporting content.


\bibliographystyle{unsrt} 
\bibliography{mainsource}

\end{document}


\maketitle



\section{Frequency transition of the USRP signal}
A perfect random access RF source would instantaneously change the frequency and amplitude of the emitted wave according to the given control parameters. The goal of the analysis presented in this section is to provide a measurement of the time that is needed to the USRP to change the frequency and amplitude of the emitted RF wave.
For this purpose, we define a model and we fit the measured USRP transmitted signal shown in Fig.3 (a) of the main article. We recall that the corresponding software set frequencies are 160 MHz and 290 MHz and that the emission duration for each RF frequency is set to 200 ns. The measured signal contains two frequency transitions that will be analyzed: 160 $\stackrel{\rightarrow}{}$ 290 MHz and 290 $\stackrel{\rightarrow}{}$160 MHz.

\subsection{Model}
The USRP transmitted signal $S(t)$ can be described as a sinusoidal function whose argument $\phi (t)$ and amplitude $a(t)$ vary in time.

\begin{equation}
S(t) = a(t) \sin{\phi(t)}
\label{eq:sig}
\end{equation}

We suppose that the USPR changes from the initial frequency $f_0$ to the final frequency $f_1$ following an exponential decay law with time constant $\tau$:
\begin{equation}
f(t) = f_0 + (f_1 - f_0)\left(1- e^{-\frac{t-t_s}{\tau}}\right)\cdot H(t-t_s).
\label{eq:freq}
\end{equation}
The time at which the frequency transition starts is called $t_s$, it represents the time at which the USRP triggers the parameters change. $H(t-t_s)$ is the unit step function (equal to 0 for $t<t_s$ and to 1 for $t>t_s$). \\From Eq.\ref{eq:freq} we derive the argument of the sinusoidal function as following: 
 
\begin{align}
\phi(t) & = 2\pi \int{f(t) \text{d}t} +\phi_0\nonumber\\
& = 2\pi f_0 t + (f_1 - f_0) \tau \left( \frac{t-t_s}{\tau} + \exp{-\frac{t-t_s}{\tau}} + C\right)\cdot H(t-t_s)  +\phi_0\\
\label{eq:phi_int}
& = 2\pi f_0 t + \Delta \phi(t) + \phi_0 \nonumber
\end{align}
where $C$ is the integration constant and $\Phi_0$ the initial phase. By imposing the continuity condition $\Delta \phi(t=t_s) = 0$, one can find $C=-1$. Thus, the expression for $\phi(t)$ is finally written as: 
\begin{equation}
\phi(t) = 2\pi f_0 t + (f_1 - f_0) \tau \left( \frac{t-t_s}{\tau} + e^{-\frac{t-t_s}{\tau}}-1\right)\cdot H(t-t_s)  +\phi_0.
\label{eq:phi}
\end{equation}

Concerning the signal amplitude, one can see in Fig.3 (a) that it oscillates for about 15 ns before stabilizing while the frequency is changed. In order to quantify the amplitude transition time more precisely, this behavior is described by supposing that the amplitude starts oscillating at the same time $t_s$ as the frequency transition and that its oscillation is damped with a time constant $\tau_a$. Therefore one can write the amplitude time evolution as:

\begin{equation}
a(t) = a_0 + (a_1 - a_0)\left( 1 + A \sin(2 \pi f_a (t-t_s) + \phi_a) e^{-\frac{t-t_s}{\tau_a}}  \right)\cdot H(t-t_s);
\label{eq:amp}
\end{equation}
where $a_0$ and $a_1$ are respectively the amplitude of the initial and final wave and $A$, $f_a$, $\phi_a$ are the amplitude, frequency and initial phase of the amplitude oscillations occurring for $t>t_s$. 
  
\subsection{Fit procedure and results}
The quantities that we are mainly interested in are $\tau$ and $\tau_a$, respectively the time the USRP takes to change frequency and amplitude of the emitted signal. To retrieve their values we realized different data fitting steps.\\ 
We first fitted the stationary oscillations at the nominal frequency of 160 MHz and 290 MHz respectively with the following functions:
\begin{align}
S_0(t) & =  a_0 \sin( 2\pi f_0 t +\phi_0) \label{eq:sin0}\\
S_1(t) & =  a_1 \sin( 2\pi f_1 t +\phi_1).
\label{eq:sin1}
\end{align}
Where with "stationary oscillations" we mean the transmitted signal away from the instant $t_s$, when amplitude and frequency are constant.The results are shown in Table \ref{tab:fit0}.
One can remark that there is a discrepancy of less than 1\% between the set frequencies of the USRP and the values of $f_0$ and $f_1$ obtained by the fit. This can be explained by the fact that the oscilloscope used for the acquisition wasn't synchronized with the USRP clock. 
\begin{table}[htbp]
\centering
\caption{\bf Fit parameters of the stationary signals $S_0$ and $S_1$ (Eqs. \ref{eq:sin0}-\ref{eq:sin1})}
\begin{tabular}{cc}
\hline
Parameter & Fit value  \\
\hline
$a_0$ & 177.3 $\pm$ 0.4 mV \\
$f_0$ &  158.37 $\pm$ 0.04 MHz \\
$\phi_0$ & 3.340 $\pm$ 0.004 rad\\
$a_1$ & 167.2 $\pm$ 0.2 mV\\
$f_1$ &  291.66 $\pm$ 0.02 MHz \\
$\phi1$ & - 1.445 $\pm$ 0.002 rad\\
\hline
\end{tabular}
  \label{tab:fit0}
\end{table}\\

\begin{figure}
    \centering
    \subfigure[]{\includegraphics[width=0.49\textwidth]{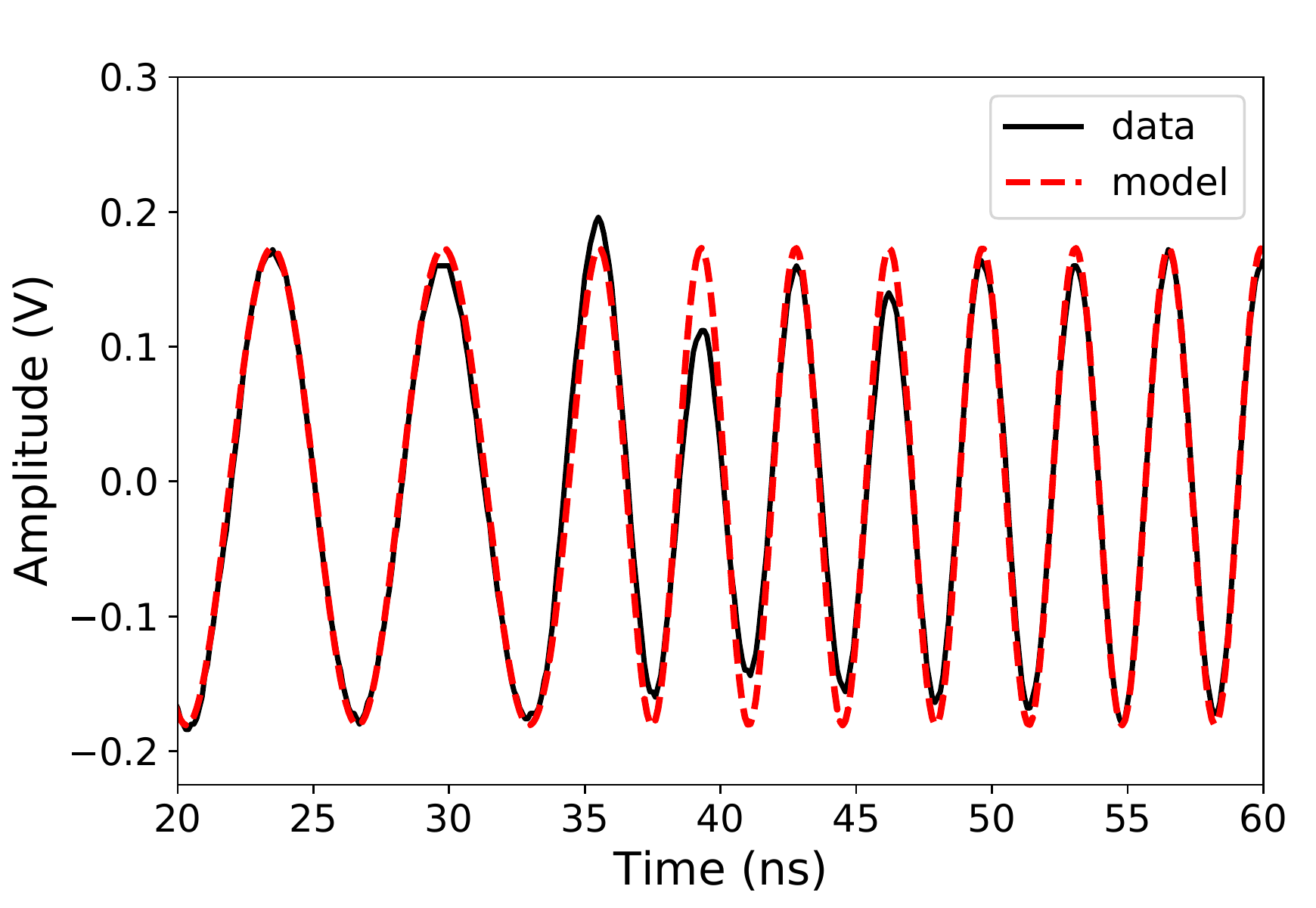}\label{fig:fit1f}} 
    \subfigure[]{\includegraphics[width=0.49\textwidth]{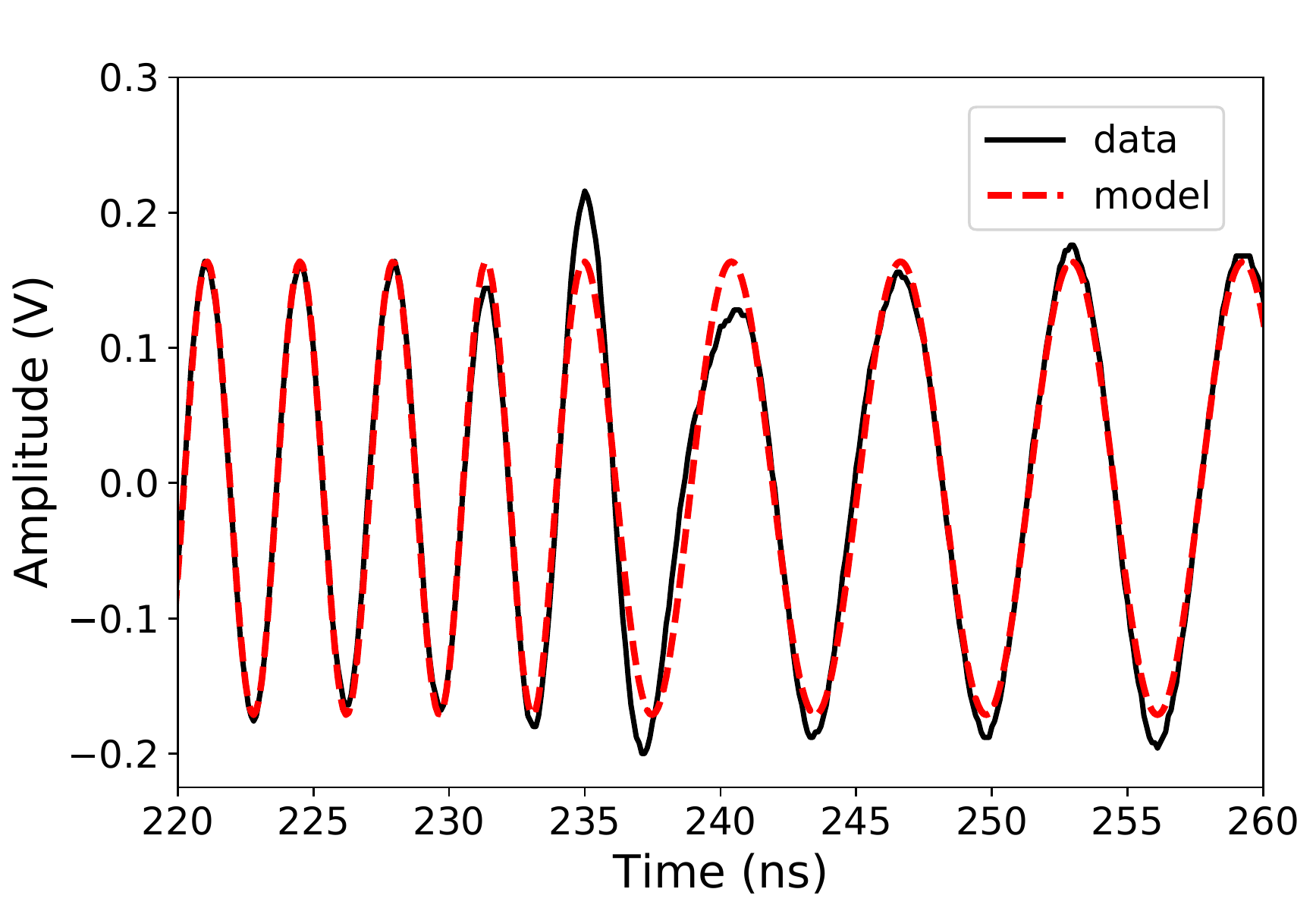}\label{fig:fit2f}} 
    \caption{Fit of the USRP transmitted signal during two frequency transitions: 160 $\stackrel{\rightarrow}{}$290 MHz (a), 290 $\stackrel{\rightarrow}{}$160 MHz (b); using model \ref{eq:sig} with \ref{eq:phi} and constant amplitude. Experimental data is shown in black straight lines and the fit in red dashed lines.}
     \label{fig:fit0}
\end{figure}

Then, the two frequency transitions (160 $\stackrel{\rightarrow}{}$ 290 MHz and 290 $\stackrel{\rightarrow}{}$160 MHz) are fitted using Eq.\ref{eq:sig} with $\phi(t)$ given by Eq.\ref{eq:phi} and considering the amplitude constant and equal to $a(t) = a_0$ from Tab.\ref{tab:fit0}. The parameters $f_0$, $\phi_0$ and $f_1$in Eq.\ref{eq:phi} are also considered constant and equal to the values given in Tab. \ref{tab:fit0} for the transition 160 $\stackrel{\rightarrow}{}$290 MHz. For the transition 290 $\stackrel{\rightarrow}{}$160 MHz, the parameters values are exchanged: $f_0 \stackrel{}{\leftrightarrow} f_1$ $\phi_0 \stackrel{}{\leftrightarrow} \phi_1$. 
The free parameters are: the switch time $t_s$ and the characteristic transition time $\tau$. The fit program uses the Levenberg-Marquardt algorithm to minimize the residuals between the experimental data and the model, it returns the value and the standard uncertainty of each fitted parameter \cite{Newville14}. The results are shown in Figs.\ref{fig:fit0} and the corresponding parameters values with the respective standard uncertainties are in Table \ref{tab:fit1}.

\begin{table}[htbp]
\centering
\caption{\bf Fit parameters of two frequency transitions of the USRP.}
\begin{tabular}{cc}
\hline
\multicolumn{2}{c}{160 $\stackrel{\rightarrow}{}$290 MHz} \\
\hline
Parameter & Fit value  \\
$t_s$ & 33.52 $\pm$ 0.05 ns\\
$\tau$ & 2.21 $\pm$ 0.05 ns\\
\hline
\end{tabular}
\quad\quad
\begin{tabular}{cc}
\hline
\multicolumn{2}{c}{290 $\stackrel{\rightarrow}{}$160 MHz} \\
\hline
Parameter & Fit value  \\
$t_s$ & 233.98 $\pm$ 0.06 ns\\
$\tau$ & 1.74 $\pm$ 0.06 ns\\
\hline
\end{tabular}
\label{tab:fit1}
\end{table}
The difference between the two $t_s$ is about 200 ns, as expected, with a relative error of 0.2\%. The frequency transition seems to be slightly faster for 290 $\stackrel{\rightarrow}{}$160 MHz. The mean value of $\tau$ is $1.98\pm0.24$ ns.\\ The fitted curves are already in good agreement with the experimental data. Differences concern mainly the signal amplitude close to the transitions time $t_s$.\\
To improve the fit, we finally consider in our model (Eq.\ref{eq:sig}) the temporal evolution of the amplitude (Eq.\ref{eq:amp}). Only the parameters concerning the amplitude variations are considered free, except the initial phase $\phi_a$ that has been constrained to satisfy the expression $\phi_a = \arcsin{-1/A}$ to guarantee the continuity of $a(t)$ at $t = t_s$ (Eq. \ref{eq:amp}). All other parameters are considered constant and equal to the values in Tables \ref{tab:fit0} and \ref{tab:fit1}. The fit results for each transition together with the parameter description are in Table \ref{tab:fit2}. The corresponding curves are shown in Fig.3(b) of the main article where on can see that the overlap of the fit with the data is improved compared to Figs. \ref{fig:fit0}.\\ We are only interested in the amplitude characteristic time $\tau_a$. It is similar in both transitions and its mean value is  $7.1 \pm 0.3$ ns. Because the model was built on the phenomenon observation and not on physical assumptions, no speculations will be done on the meaning of the other parameters.   

\begin{table}[htbp]
\centering
\caption{\bf Fit parameters of the amplitude's temporal evolution of the signal emitted by the USRP during the two frequency transitions.}
\begin{tabular}{l}

\\
\hline
Parameter description\\
\hline
Modulation amplitude\\
Modulation frequency\\
Damping time\\
\hline
\end{tabular}
\quad\quad
\begin{tabular}{cc}
\hline
\multicolumn{2}{c}{160 $\stackrel{\rightarrow}{}$290 MHz} \\
\hline
Parameter & Fit value  \\

$A$ & -5.5 $\pm$ 0.4\\
$f_a$ & 129 $\pm$ 2 MHz\\

$\tau_a$ & 7.4 $\pm$ 0.7 ns\\
\hline
\end{tabular}
\quad\quad
\begin{tabular}{cc}
\hline
\multicolumn{2}{c}{290 $\stackrel{\rightarrow}{}$160 MHz} \\
\hline
Parameter & Fit value  \\

$A$ & 9.0 $\pm$ 0.4
\\
$f_a$ & 143 $\pm$ 1
MHz\\

$\tau_a$ & 6.8 $\pm$ 0.4
ns\\
\hline
\end{tabular}
\label{tab:fit2}
\end{table}



\section{Deflection speed}
In Sec. 4.2 of the main article, the transition time between two random positions was measured by considering the 90\% of the signal level on two photodiodes (PDs). Moreover it was observed that the PDs' signals oscillate before stabilizing (see Fig.\ref{fig:PDs} that is a copy of Fig. 5(c) of the main article). The signal of PD1 goes up first and then stabilizes at a lower level (red dots in Fig. \ref{fig:PDs}) while the signal of PD2 starts below the final level (black crosses). This phenomenon does not depend on the photodiode used for detection and in this work we aim at describing it more precisely. In the following we will fit the experimental data of each PD's signal with the step response of a second order filter.
\subsection{Model}
A second order filter is the simplest system that can show an oscillatory behavior in response to a step input. Such response to a unitary step is described by the following analytical formula \cite{Kuo2010}:
\begin{equation}
P(t) = \left( 1 - \sin{\left(2 \pi \nu_a \sqrt{1-\xi^2} (t-t_0) + \arccos{\xi}\right)} \frac{e^{- 2 \pi \nu_a \xi(t-t_0)}}{\sqrt{1-\xi^2}}  \right)\cdot H(t-t_0);
\label{eq:pds}
\end{equation}
where $\nu_a$ is the resonant frequency, $t_0$ is the instant at which the unit step input is given and $\xi$ is the damping ratio. Eq.\ref{eq:pds} is valid only for the under damped case i.e. for $\xi$ < 1.  
\subsection{Fit procedure and results}
\begin{figure}[htbp]
\centering\includegraphics[width=1\linewidth]{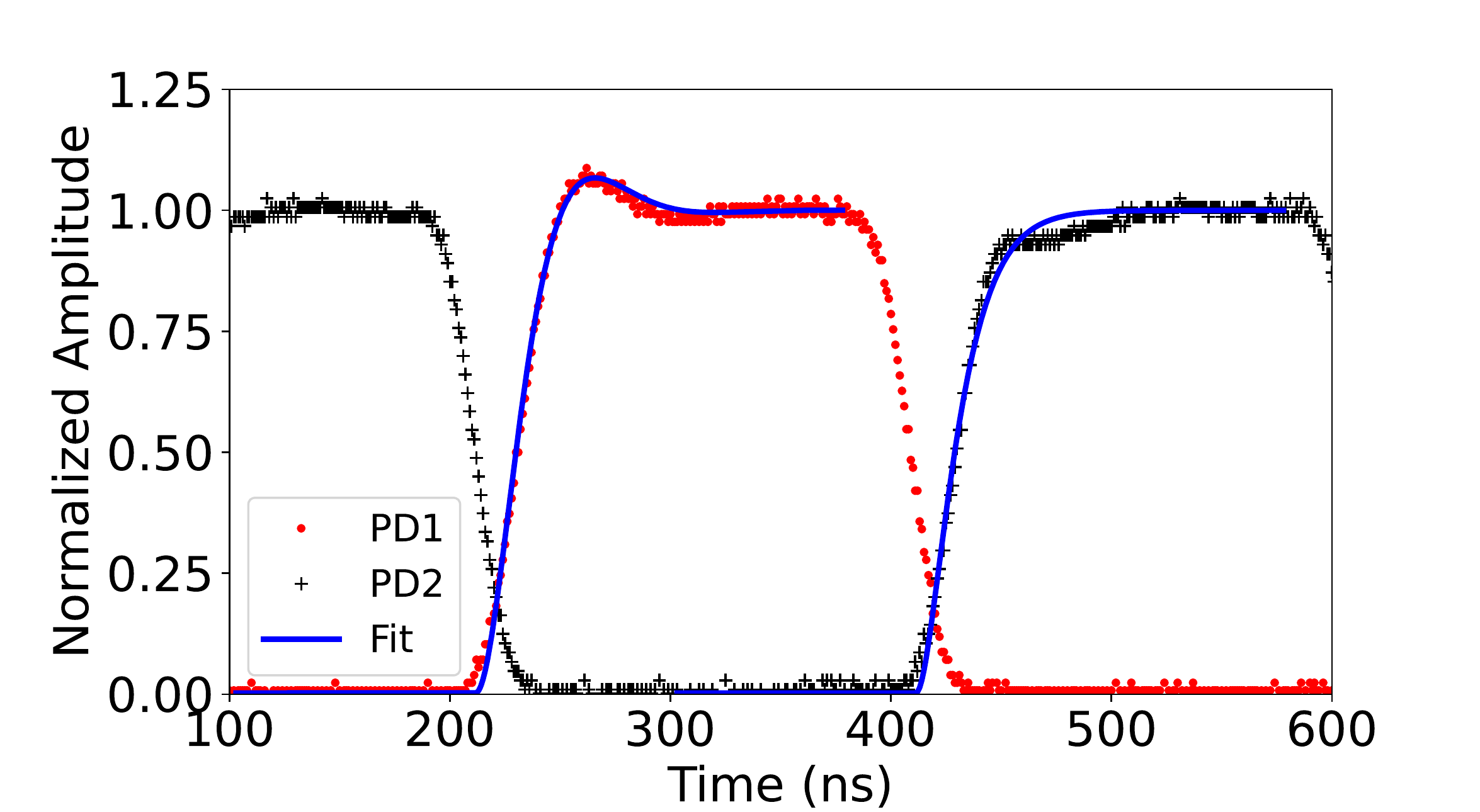}
\caption{Normalized amplitude of the photodiodes signals: PD1 (red dots), PD2 (black crosses). And fit of their rising signal. Each photodiode corresponds to a different beam transverse position (see main article).  }
\label{fig:PDs}
\end{figure}

We fit the model to the normalized signals of PD1 and PD2 (see Fig. \ref{fig:PDs}). The corresponding parameters are in in Tab. \ref{tab:fitPDs}.\\ 
One can see that our model is able to describe the temporal evolution of the PD1 signal quite well. The chi-squared value given by the fit algorithm is 0.06. 
The rising signal of PD2 is also well described by the model but the deviation with respect to the data is more important compared to PD1. In this case the chi-squared value is 0.12. Moreover the damping ratio is $\xi\approx 1$, that is the validity limit of Eq.\ref{eq:pds}. Therefore we also fitted this curve with the step response solution of a over-damped system ($\xi$>1) and we obtained again a value very close to unity: $\xi = 1.03 \pm 0.04$, meaning that the system is better described by a critically damped system ($\xi = 1$). A fit with the response of a critically damped system was also performed and gave the same values, within the error bars, for $t_0$ and $\nu_a$. The goodness of the fit was also identical so the first fit has been retained.\\ 

\begin{table}[htbp]
\centering
\caption{\bf Fit parameters of the rising signal $P(t)$ of the two photodiodes (PD1 and PD2).}

\quad\quad
\begin{tabular}{cc}
\hline
\multicolumn{2}{c}{PD1} \\
\hline
Parameter & Fit value  \\

$t_0$ & 211.7 $\pm$ 0.2 ns
\\
$\nu_a$ & 12.2 $\pm$ 0.1 MHz
\\
$\xi$ & 0.652 $\pm$ 0.006\\
\hline
\end{tabular}
\quad\quad
\begin{tabular}{cc}
\hline
\multicolumn{2}{c}{PD2} \\
\hline
Parameter & Fit value  \\

$t_0$ & 411.4 $\pm$ 0.2 ns
\\
$\nu_a$ & 15.2 $\pm$ 0.4 MHz
\\
$\xi$  & 0.99 $\pm$ 0.01
\\

\hline
\end{tabular}
\label{tab:fitPDs}
\end{table}
Our goal is to measure the rise time, an indicator of the deflection speed. According to \cite{Levine2011}, the rise time of an under-damped second order system is the time required to go from 0\% to 100\% of its final value, right after the overshoot. So the rise time $\tau_r$ is obtained by setting the sine argument of Eq.\ref{eq:pds} equal to $\pi$:
\begin{equation}
\tau_r = \frac{\pi - \arccos{\xi}}{2\pi\nu_a\sqrt(1-\xi^2)}.
\label{eq:tr}
\end{equation}
For the PD1 signal $\tau_r = 39 \pm 2$ ns, where the uncertainty has been calculated from the parameters' standard deviation in Tab. \ref{tab:fitPDs} using the error propagation. Always according to \cite{Levine2011}, the rise time for a critically damped system is the time required to rise from 5\% to 95\% of its final value. Therefore, for the PD2 signal the rise time is 45 ns when measured on the fitted curve and 42 ns when measured on the data.\\ 

We recall that the PD1 (PD2) signal corresponds to the response of the 2D deflection system described in the main article, when the input frequencies are changed from 290(160) MHz to 160(290) MHz. These measurements confirm that the rapidity of our deflection system is limited by the AOM's response which was expected of the order of tens of ns\footnote{The rise time given by the manufacturer is 33 ns without clear definition of rise time. This value is given for a beam diameter of 210 $\mu$m while our measurement is realized with a beam diameter of 190 $\mu$m.} .The USRP output instead evolves on a shorter time scale, its longer characteristic time is 7.1 ns.
This value can be compared more directly with the exponential time constant $1/(2\pi\nu_a\xi) = 20$ ns that describes the damping of the oscillations in Eq. \ref{eq:pds}.\\




\bibliography{supplementsource}
